\documentclass[twocolumn,superscriptaddress,amssymb,amsmath,floatfix,citeautoscript]{revtex4-1}

\usepackage[pdftex]{graphicx}
\usepackage{chngcntr}
\usepackage{amsmath}
\usepackage{color}
\usepackage{natbib}
\usepackage{mathtools}
\usepackage{physics}
\allowdisplaybreaks

\newcommand{\rmf}{Rb$_2$MnF$_4$}

\begin{document}
%
%
%
%
%
%--------------------------------------------------------------------- %
\title{Implementation of a laser-neutron pump-probe capability at HYSPEC}
%--------------------------------------------------------------------- %

\author{C. Hua}
\email{ huac@ornl.gov}
\affiliation{Materials Science and Technology Division, Oak Ridge National Laboratory, Oak Ridge TN, 37831, USA}
\author{D.A. Tennant}
\affiliation{Department of Physics and Astronomy, University of Tennessee, Knoxville, TN 37996, USA }
\affiliation{Department of Materials Science and Engineering, University of Tennessee, Knoxville TN, 37996, USA}
\affiliation{Shull Wollan Center, Oak Ridge National Laboratory, Oak Ridge TN, 37831, USA }
\author{A. T. Savici}
\affiliation{Neutron Scattering Division, Oak Ridge National Laboratory, Oak Ridge TN, 37831, USA}
\author{V. Sedov}
\affiliation{Neutron Scattering Division, Oak Ridge National Laboratory, Oak Ridge TN, 37831, USA}
\author{G. Sala}
\affiliation{Neutron Scattering Division, Oak Ridge National Laboratory, Oak Ridge TN, 37831, USA}
\author{B. Winn}
\email{winnbl@ornl.gov}
\affiliation{Neutron Scattering Division, Oak Ridge National Laboratory, Oak Ridge TN, 37831, USA}

\date{\today}

\widetext

This work was supported by the U. S. Department of Energy, Office of Science, Basic Energy Sciences, Materials Sciences and Engineering Division and National Quantum Information Science Research Centers, Quantum Science Center. This manuscript has been authored by employees of UT-Battelle, LLC under Contract No.~DE-AC05-00OR22725 with the U.S. Department of Energy. The publisher, by accepting the article for publication, acknowledges that the U.S. Government retains a nonexclusive, paid-up, irrevocable, worldwide license to publish or reproduce the published form of this manuscript, or allow others to do so, for U.S. Government purposes. This research used resources at the Spallation Neutron Source, a DOE Office of Science User Facility operated by the Oak Ridge National Laboratory. 

\thispagestyle{empty}

\newpage

\clearpage
\pagenumbering{arabic} 

\begin{abstract}

Exciting new fundamental scientific questions are currently being raised regarding nonequilibrium dynamics in spin systems, as this directly relates to low power and low loss energy transport for spintronics. Inelastic neutron scattering (INS) is an indispensable tool to study spin excitations in complex magnetic materials. However, conventional INS spectrometers currently only perform steady-state measurements and probe averaged properties over many collision events between spin excitations in thermodynamic equilibrium, while the exact picture of re-equilibration of these excitations remains unknown. In this work, we designed and implemented a time-resolved laser-neutron pump-probe capability at HYSPEC (Hybrid Spectrometer, beamline 14-B) at the Spallation Neutron Source (SNS) at Oak Ridge National Laboratory. This capability allows us to excite out-of-equilibrium magnons with a nanosecond pulsed laser source and probe the resulting dynamics using INS. Here, we discussed technical aspects to implement such a capability in a neutron beamline, including choices of suitable neutron instrumentation and material systems, laser excitation scheme, experimental configurations, and relevant firmware and software development to allow for time-synchronized pump-probe measurements. We demonstrated that the laser-induced nonequilibrium structural factor is able to be resolved by INS in a quantum magnet. The method developed in this work will provide SNS with advanced capabilities for performing out-of-equilibrium measurements, opening up an entirely new research direction to study out-of-equilibrium phenomena using neutrons.

\end{abstract}

\maketitle

\section{Introduction}

Neutron scattering is a unique technique for understanding magnetism in condensed matter, because the neutron carries a spin, and therefore a magnetic moment, that can couple directly to different sources of magnetism in matter such as unpaired electrons. In fact, uncovering antiferromagnetism was among the very first experiments utilizing neutron diffraction.\cite{shull_detection_1949} Since then, neutron scattering has proven to be a powerful tool for the investigation of magnetic structures. With the invention of triple-axis neutron spectrometer and time-of-flight direct geometry spectrometers (DGSs), inelastic neutron scattering (INS) has become an indispensable tool to study momentum-resolved spin excitations in increasingly complex magnetic materials.\cite{bayrakci_lifetimes_2013, dalla_piazza_fractional_2015,muhlbauer_skyrmion_2009,scheie_detection_2021}

Recently, nonequilibrium dynamics of excited spin states in quantum magnets has gained great attention because it promises transformative technologies such as high speed, low power and low loss electronics.\cite{wolf_spintronics_2001, noauthor_rise_2016, noauthor_upping_2018, jungwirth_multiple_2018,flicker_understanding_2022, hallen_dynamical_2022, scheie_detection_2021} Once quasiparticles such as magnons (quantized spin waves) in quantum magnets are excited to out-of-equilibrium states , they remain out-of-equilibrium for a long time due to their restrictive scattering phase space and a weak coupling with phonons, creating a dynamical bottleneck. However, re-equilibration and thermalization of nonequilibrium magnon states in quantum magnets is not well understood and mode-dependent information of energy flows among magnons during re-equilibration is unknown. Conventional INS spectrometers do not elucidate such information because they currently only perform steady-state measurements at thermodynamic equilibrium and probe averaged properties of many collision events between quasiparticles. Because neutron scattering is such an important tool to provide momentum resolved information about the underlying quasiparticles, it is necessary to develop a new experimental platform and methodology for uncovering hidden physics of out-of-equilibrium excited states using neutrons.   

Time-resolved pump-probe techniques based on optical spectroscopy have been used for decades to study out-of-equilibrium dynamics.\cite{fischer_invited_2016} Pump-probe-type experiments at the user facilities such as the Advanced Photon Source \cite{dufresne_time-resolved_2010} and Linac Coherent Light Source \cite{glownia_pumpprobe_2019} have enabled scientific investigations into the ultrafast time-resolved evolution of atoms, molecules, bio-systems and solid-state systems.\cite{faatz_development_2001, teitelbaum_direct_2018,cammarata_impulsive_2006,stoica_optical_2019,epp_time_2017, wall_ultrafast_2018} In a typical pump-probe experiment, an optical laser pulse initiates dynamics that are later probed by an X-ray pulse.\cite{bionta_spectral_2011,hartmann_sub-femtosecond_2014} The X-ray probe measurement can employ a wide variety of methods including scattering, diffraction, emission, X-ray near-edge absorption, \emph{etc.} to capture the dynamics.

\begin{figure}
\includegraphics[scale = 0.25]{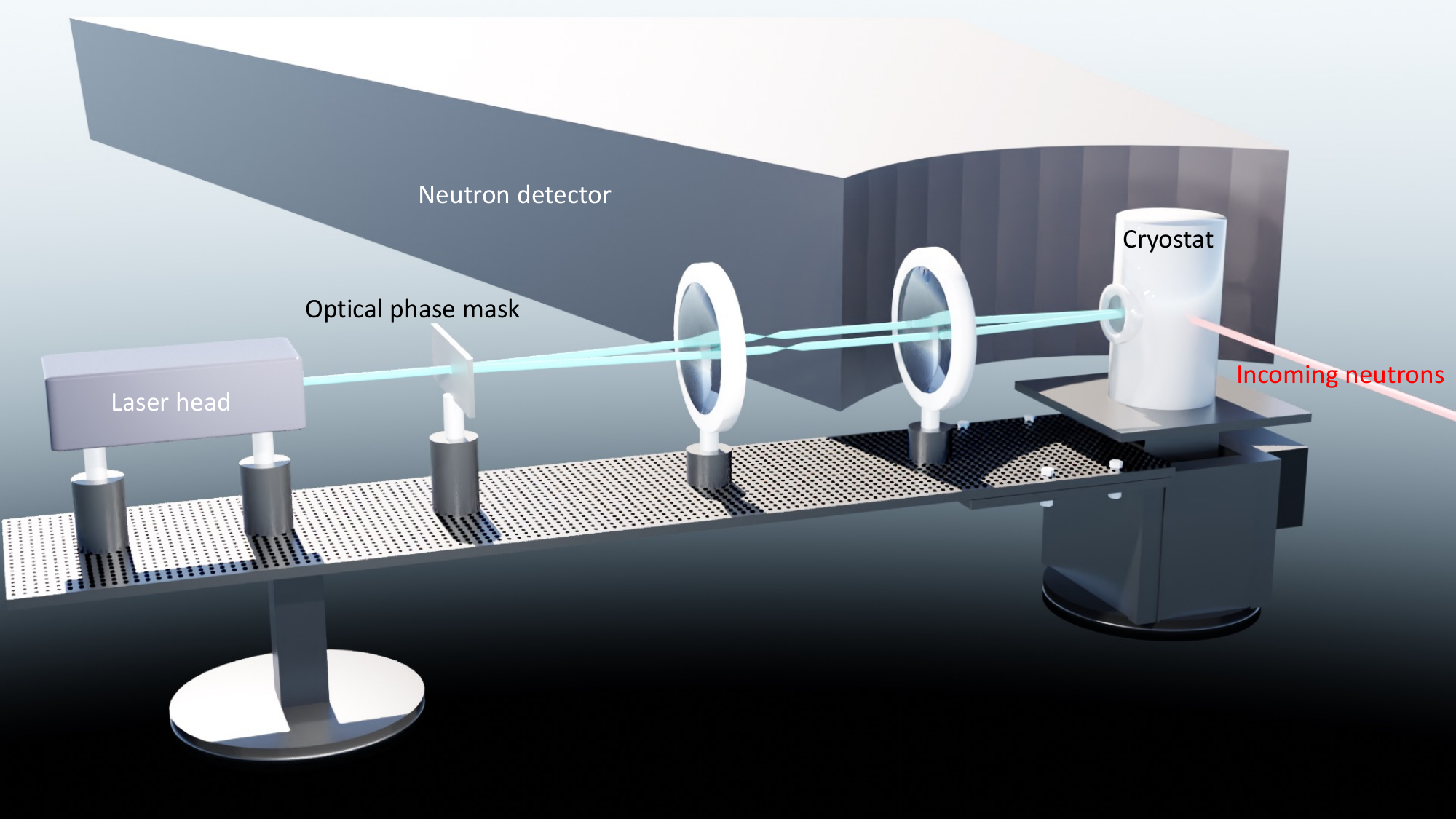}
\caption{Conceptual schematic of a laser-pump neutron-probe technique. The laser setup is attached to the sample rotation stage using a level system, allowing the laser enclosure to rotate with a neutron-compatible optical cryostat and always ensuring normal incidence of the laser beam onto the sample during the INS measurements.}
\label{fig:SetupSchematics}
\end{figure}

Time-resolved pump-probe neutron experiments, on the other hand, are much more challenging to perform with fewer works reported. This is because much fewer neutron sources worldwide compared to X-ray facilities limit the development of novel sample environments and new techniques in a beamline, and the low neutron flux that a typical beamline can produce usually limits the achievable time-resolution which in turn defines the scope of dynamics one can investigate. Most time-resolved neutron measurements in the past have focused on soft matter.\cite{urban_soft_2021}

Earliest works that enabled time-resolved neutron capability dated back to the early 1990s, where a trigger is used to initiate a non-reversible process in polymers and the data are collected in individual time frames with time resolutions of a few minutes per frame using small angle neutron scattering (SANS) techniques.\cite{jinnai_time-resolved_1991} As neutron instrumentation advanced over the last three decades, the achievable time resolution has drastically improved from minutes per frame to sub-millisecond per frame.\cite{connell_phase-separation_1991,niimura_aggregation_1995,egelhaaf_micelle--vesicle_1999,mason_time-resolved_2003,wiedenmann_dynamics_2006,hollamby_growth_2012,mahabir_growth_2013} Instrumentation for time-resolved SANS measurements with sub-millisecond time resolution, based on G\"{a}hler's TISANE (time-involved small-angle neutron experiments) concepts,\cite{kipping_small_2008} has been implemented at a few neutron centers worldwide.\cite{glinka_sub-millisecond_2020,wiedenmann_dynamics_2006,muhlbauer_intrinsic_2009} This technique uses novel electronics for synchronizing the neutron pulses from high-speed counter rotating choppers with a periodic stimulus applied to a sample.\cite{adlmann_towards_2015} Besides SANS, laser-neutron pump-probe experiments have been reported to investigate light-induced protein dynamics using quasielastic neutron scattering (QENS).\cite{pieper_transient_2008, pieper_protein_2009} Neutron reflectometry in time-of-flight mode has been used to study time-resolved \emph{in situ} rheology in a micellar solution undergoing shear thinning.\cite{adlmann_towards_2015} Very recently, optically induced steady-state magnetization at perovskite/ferromagnetic interface was probed by a polarized neutron reflectometry experiment with \emph{in situ} photoexcitation at room temperature.\cite{wang_optically_2021} Time-dependent magnetic neutron diffraction  has been reported to probe magnetic dynamics of a spin ice driven by a periodic pulsed magnetic field.\cite{wang_monopolar_2021}

To study out-of-equilibrium dynamics of quasiparticles like magnons in quantum magnets, a time-resolved pump-probe technique using INS at cryogenic temperatures is necessary but has yet been attempted. In this work, we designed and implemented a pump-probe capability at HYSPEC (Hybrid Spectrometer, beamline 14-B at the Spallation Neutron Sources at Oak Ridge National Laboratory)\cite{zaliznyak_polarized_2017} that excites out-of-equilibrium magnons with a nanosecond pulsed laser source and probes the resulting dynamics using INS. A conceptual schematic of the experimental configuration is shown in Fig.~\ref{fig:SetupSchematics}. The laser setup is attached to the sample rotation stage, allowing the laser enclosure to rotate with a neutron-compatible optical cryostat and always ensuring normal incidence of the laser beam onto the sample during the INS measurements. To successfully probe the out-of-equilibrium dynamics of magnons using INS, it is necessary to have (1) an external excitation source that can create an out-of-equilibrium distribution of spin excitation states in the bulk of a crystal, (2) a material system with out-of-equilibrium dynamics that are sufficiently long lived, and (3) a pump that can be synchronized to the neutron source. Each constraint poses technical challenges that are unique to each material system of interest. In the following sections we describe technical aspects to implement such a capability in a neutron beamline, including choices of suitable neutron instrumentation and material systems, laser excitation scheme, experimental configurations, and relevant firmware and software development to allow for time-synchronized pump-probe measurements. We then present the intensity map of the laser induced non-equilibrium structural factor, $I(Q,\omega)$, measured by INS at HYSPEC in a quantum magnet and conclude with an overview of the future studies that could benefit from this type of laser-neutron pump-probe experiment as next generation neutron sources come online within the next decade. 
\section{Instrument and Material choices} 

\begin{figure}
\includegraphics[scale = 0.3]{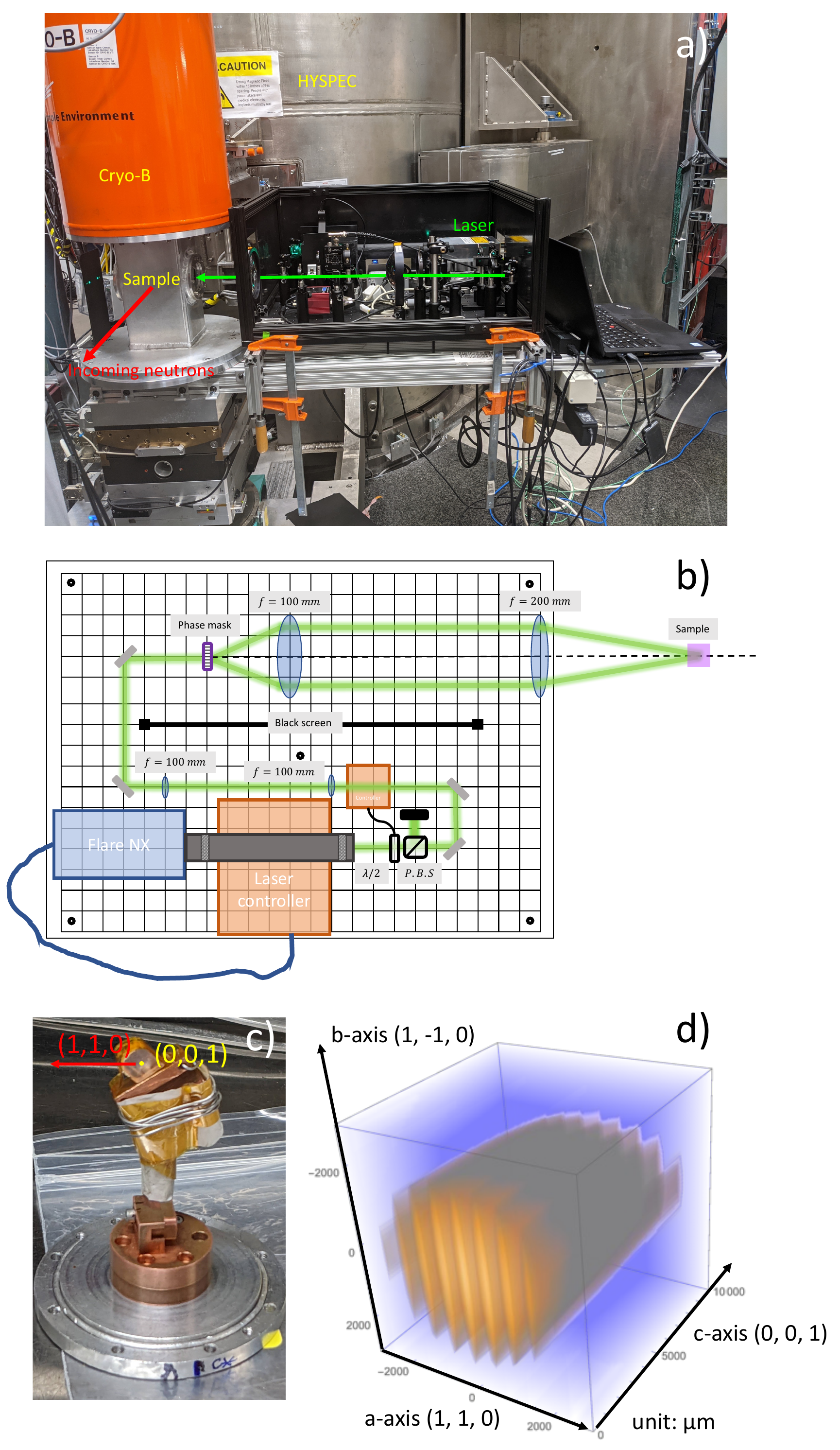}
\caption{(a) Picture of the laser setup integrated into HYSPEC beamline with Cryo-B installed. (b) Schematic of optical layout. Two time-coincident, coherent pump pulses crossed at the sample position to form a spatial interference pattern with an adjustable periodicity from 1 to 30 $\mu$m. (c) \rmf ~crystal used in this study on a copper sample mount. (d) Periodic grating pattern caused by interference of $\pm1$ order diffracted laser beams crossed at the sample.}
\label{fig:SetupPicture}
\end{figure}

HYSPEC is a high-intensity, medium-resolution, cold to thermal direct DGS and employs a hybrid design by adding the pre-sample Bragg optics found on a triple-axis spectrometer to a traditional time-of-flight DGS. HYSPEC has been routinely utilized to measure spin excitations in quantum magnets at equilibrium conditions (see e.g. figure 4 of Ref.\cite{landolt_spin_2022}). There are four main factors that make HYSPEC a suitable instrument for probing out-of-equilibrium magnon dynamics. (1) As a DGS, the pulsed peak intensity is compatible with pump-probe measurements as described later.  (2) It is optimized for inelastic neutron scattering measurements of excitations in small single-crystal specimens. The Bragg optics allow the use of vertical focusing prior to the sample position to increase the flux on the sample. (3) A $60^\circ$ range of horizontal scatter angles are accessible simultaneously via a wide-angle detector array at HYSPEC, and this range is adjustable using a motorized stage which rotates the detector vessel about the sample position. The wide-angle detection allows a fixed position of detector vessel with access to a wide region of reciprocal space in a crystal. (4) The open space available around the sample stage allows the implementation of a laser setup. As shown in Fig.\ref{fig:SetupPicture}(a), an optical setup on a 24''x18'' Thorlabs\textsuperscript{\textregistered} aluminum breadboard is mounted on 2'' rails (80/20 Inc.\textsuperscript{\textregistered}) which are attached to the sample rotation stage. 

The "orange" cryostat (CRYO-B) supplied by A.S. Scientific\textsuperscript{\textregistered} is a neutron-compatible optical cryostat with two 2'' quartz windows on the outer vacuum chamber and two 70 mm diameter sapphire windows on the inner vacuum chamber, and requires continuous flow of liquid Helium to reach a base temperature of 3.4 K. Its aluminum heat shield has rectangular sections cut out to allow the laser light to pass through both the sample and the entire cryostat.  The inner vacuum chamber at beam elevation is machined from a solid aluminum rectangular block and all quartz windows include a stainless steel frame and fasteners.  CRYO-B is normally utilized with the incident beam passing through the quartz windows, but in this configuration the incident neutron beam passes through the side walls and for some orientations through the stainless window frame and fasteners, leading to significant neutron scattering which introduces a background that varies both with scatter angle and with the orientation of the cryostat.  This background is mitigated partly by radial collimation and shielding about the sample, but must still be corrected for in reduction by subtracting the scattering signal obtained with the sample removed. 

Figure \ref{fig:SetupPicture}(c) shows the sample used in this study. The sample size is about 10 x 8 x 3 mm, dimensions chosen to be compatible with the achievable laser beam size as discussed in the next section. The composition of the sample is \rmf, a spin-5/2 Heisenberg antiferromagnet. \rmf~crystal is orthorhombic with $a=b=4.28 \AA{}$ and $c=14 \AA{}$ and magnon transport is confined to $ab$ plane. The crystal has an optical penetration depth of tens of centimeters in the visible light range and thus permits a transmission configuration of the visible light. This is an important factor to consider in a laser-neutron pump-probe experiment. INS at HYSPEC is sensitive to the bulk of a sample and thus any external stimuli needs to induce a bulk excitation. A material with a centimeter-long optical penetration depth will allow the laser to excite the bulk of the crystal rather than just induce a surface effect. Besides its physical appearance, \rmf~ provides a great platform to understand nonequilibrium magnon dynamics.  As a model Heisenberg antiferromagnet, its magnetism and Hamiltonian were well characterized at thermodynamic equilibrium using neutrons.\cite{huberman_study_2008} High-resolution neutron resonance spectroscopy suggests magnons have very long lifetimes at low temperatures, limited only by magnetic grain boundaries.\cite{bayrakci_lifetimes_2013} A comprehensive theoretical description based on the Dyson-Maleev boson formulation predicts that magnon-phonon interactions are negligible and magnon-magnon collisions are restricted to two-in two-out processes, which creates a situation where magnons at the magnetic zone center become very long lived, up to milliseconds, due to the restricted scattering phase space.\cite{PhysRevB.3.961} HYSPEC provides neutron pulses with a pulse width of around 60 $\mu$s depending on the instrument setting, which ultimately limits the time resolution of the measurements. Non-equilibrium magnon dynamics lasting up to milliseconds, in principle, will be resolvable by neutrons. Moreover, a spin moment of 5/2 makes \rmf~ a very strong neutron scatterer with respect to magnons.\cite{huberman_study_2008,bayrakci_lifetimes_2013} The large signal-to-noise ratio will increase our chances to identify any difference in the measured $S(Q,\omega)$ under laser excitation from that measured at equilibrium condition. 
\section{Pulsed laser integration}

To excite the magnons to out-of-equilibrium states, we use a pulsed laser (Coherent Flare NX\textsuperscript{\textregistered}, Class 4) that operates at 515 nm ($\sim$ 2.4 eV) with a repetition frequency up to 2000 Hz, a pulse width of around 5 ns, and pulse energy of a 300 $\mu$J. The intense pulsed laser induces an impulsive change in mangetic anisotropy through interband excitation of hot electrons in a magnetic material and charge transfer occurs among magnetic ions, leading to magnetic excited states.\cite{hashimoto_photoinduced_2008,bliokh_spinorbit_2015,hansteen_femtosecond_2005} The laser can be externally triggered by TTL signals. The optical layout is shown in Fig.~\ref{fig:SetupPicture}(b). Right after the laser head is a motorized power control unit made of a rotating halfwave plate and cubic polarizing beam splitter (PBS). This unit enables a remote control of the laser power delivered to the sample. After the PBS, the light is horizontally polarized and two 1'' lens with focus lengths of 100 mm are used to collimate the laser beam. The last part of the optical train is adapted from a transient grating (TG) technique,\cite{wang_diffusion_2019,dennett_time-resolved_2017,short_applications_2015,vega-flick_thermal_2016} where two time-coincident coherent laser pulses created by a grating mask are crossed at the sample to form an interference pattern with period $L$. The interference pattern creates spatially periodic photoexcitation on the sample with $L$ controllable from 1 $\mu$m to 30 $\mu$m, thus providing control of the length scales of external excitations. The periodic grating pattern caused by interference of $\pm1$ order diffracted laser beams crossed at the sample, shown in Fig.~\ref{fig:SetupPicture}(d), is along the $a$-axis of the crystal. 

\begin{figure}
\includegraphics[scale = 0.4]{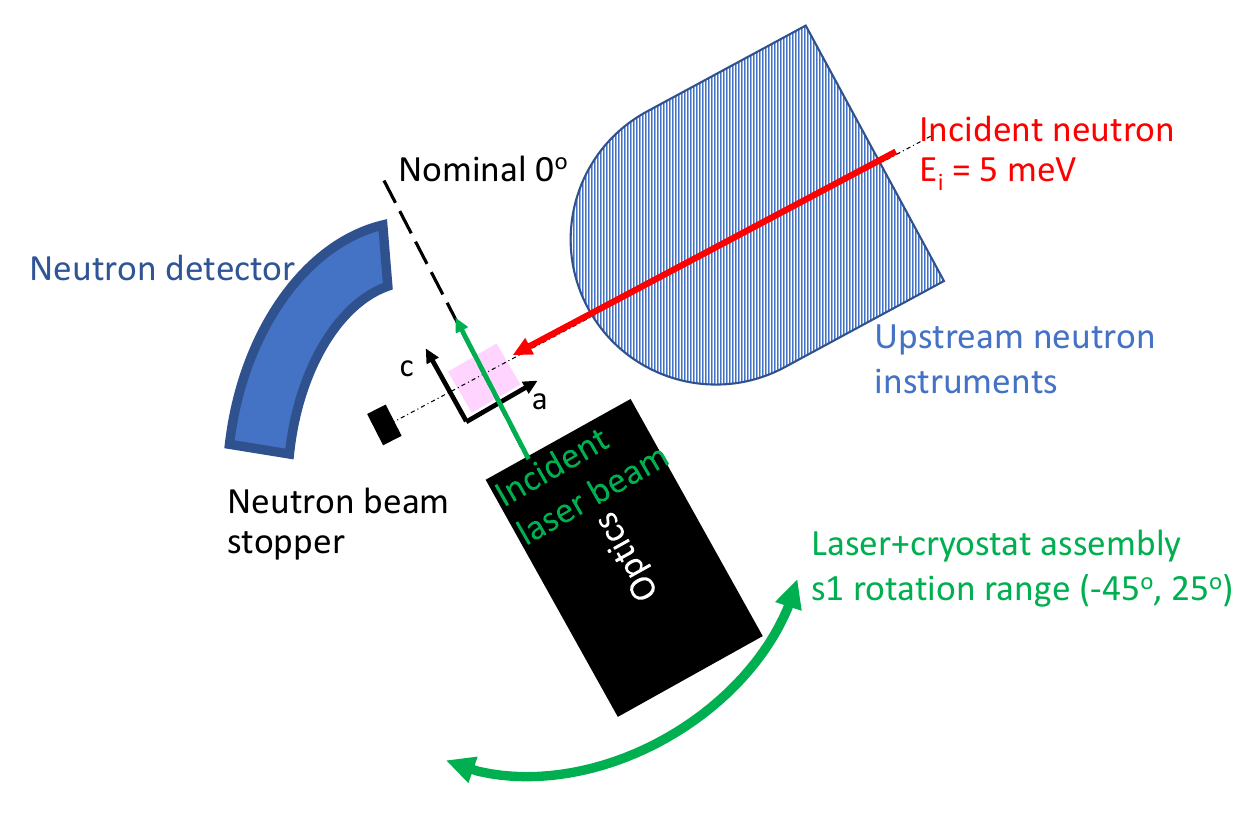}
\caption{Schematics of the experimental configuration. The rotation angle, $s_1$, of the laser+cryostat assembly is restricted to $-45^\circ < s_1 < +25^\circ$ to avoid interference of the laser enclosure with neutron beams, upstream neutron instruments, and neutron detector vessel.}
\label{fig:ExpConfig}
\end{figure}

Two 4'' lens are used here to accommodate the large diffraction angle of the smallest grating on the phase mask. The 1:2 ratio of focus lengths is used such that laser beam is enlarged to a diameter of 5 mm at the sample position. The beam size at the sample position is determined by two main factors: (1) the minimum fluence that is required to excite nonequilibrium population of magnons and (2) the minimum sample size to obtain a total structure factor with good statistics in a reasonable time frame. Here the maximum fluence delivered to the sample is 15 J/cm$^2$. The light absorption rate of \rmf~ at 515 nm is estimated to be 10\% and about 25\% of the crystal volume is illuminated by the laser. Since the laser enclosure (black box in the center of Fig.~\ref{fig:SetupPicture}(a)) is attached to the sample rotation stage, the enclosure rotates with cryo-B and, therefore, optical axis indicated as dashed line in Fig.~\ref{fig:ExpConfig} is always orthogonal to $ab$ plane of the sample during the INS measurements.

There are three important advantages of a TG-type excitation when compared to just using a Gaussian-profiled laser excitation: (1) TG is ideal for creating local non-equilibrium processes due to its controllable excitation patterns in space. The periodicity of the interference pattern varies from tens of micrometers to one micrometer, which are comparable to the mean free paths (MFPs) of magnons in quantum magnets. The nanosecond pulsed excitation is comparable to the magnon lifetimes in quantum magnets. When the external excitation length or time scales are on the order of magnons' MFPs or lifetimes, then we create a situation such that magnon transport becomes non-diffusive. Magnons with MFPs larger than the external excitation length scale or lifetimes longer than the excitation duration do not scatter within the excitation length or time scales. Thus, they remain in their nonequilibrium excited states.  (2) The TG geometry is ideal for studying spin transport parallel to a plane since the temperature gradient imposed through the interference pattern can be easily aligned with a particular direction. (3) Due to its non-contact and non-invasive nature, the scheme can be integrated in a variety of sample environments. If out-of-equilibrium magnon states are indeed long lived in \rmf ~up to 1 ms, then a neutron probe with a pulse width of 60 $\mu$s provides a time structure which can directly observe the out-of-equilibrium dynamics of magnons under such an excitation scheme. 

\begin{figure*}
\includegraphics[scale = 0.6]{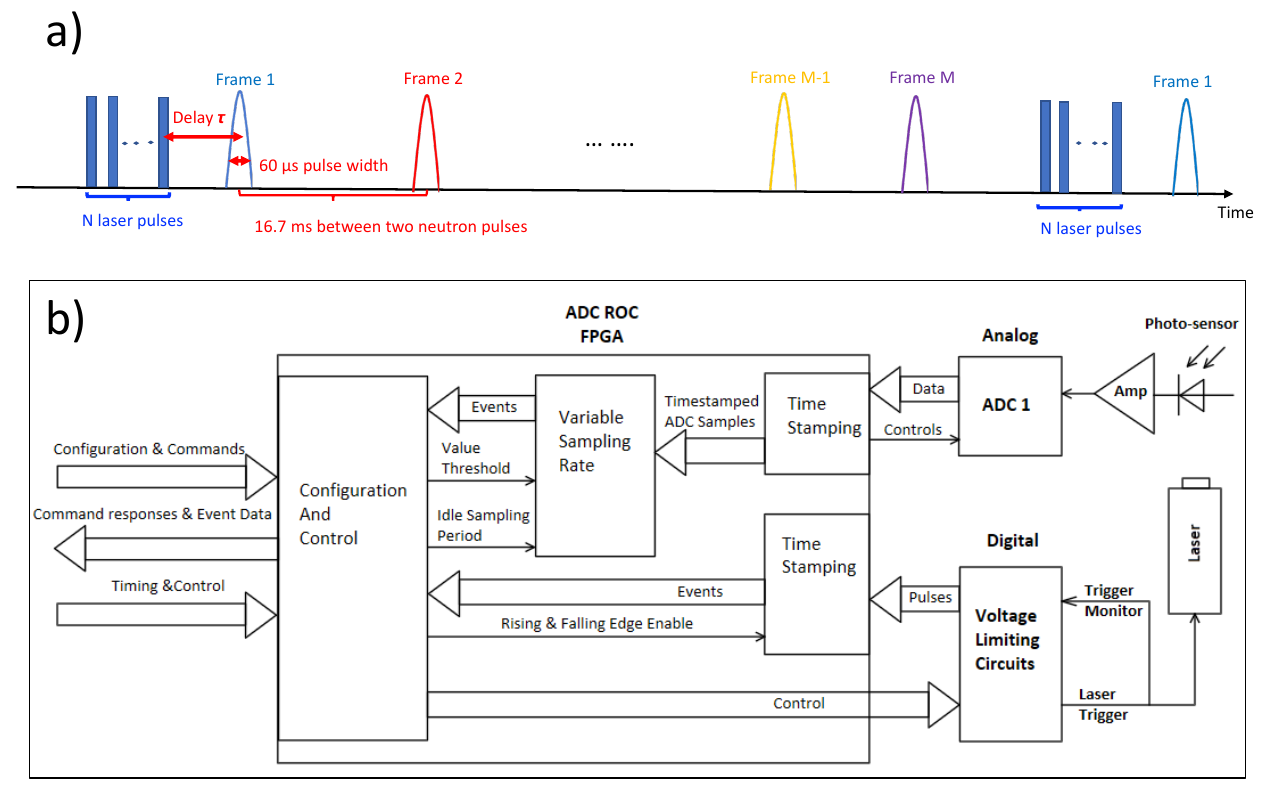}
\caption{(a) Schematics of a sequence of laser pulses are triggered by an external signal that is synchronized to the neutron spallation events with a tunable repetition frequency and delay time. (b) Diagram of the control and data acquisition setup in this experiment. One digital channel is configured as output to trigger the laser, one input analog channel is used to acquire the laser trigger signal and two digital channels were configured as inputs to acquire the accelerator time reference and laser trigger reference signals. The relative difference between the accelerator time and laser trigger references sets the delay between laser and neutron pulses arriving at the sample. In order to reduce amount of data acquired by ADCROC, a variable sampling rate functionality is utilized. The data acquisition rate increases when it detects a fast change in signals.}
\label{fig:firmware}
\end{figure*}

\section{Experimental configurations}

The requirement of suitable neutron incident and scattering angles with compatible laser incident angle onto the sample introduces a set of constraints on the experimental configuration. Schematics of the experimental configuration are shown in Fig.~\ref{fig:ExpConfig}. 

Magnon excitations in \rmf~crystal are observed about the $[0.5, 0.5, L]$ trajectory for all $L$ in lattice reciprocal space. Although magnons are also observable about $[0,0,L]$ and $[1,1,L]$, those regions are avoided due to the overlapping contribution from phonons. The HYSPEC spectrometer has an incident beam in the horizontal plane and scatters horizontally, with only limited out-of-plane divergence.  For this experiment, in order to obtain spectral information resolvable in reciprocal space in the $[H,H,L]$ plane, the crystal must be rotated about the vertical axis parallel to the $[H,-H,L]$ direction. For magnetic scattering with neutrons, the greatest signal is generally found at the lowest momentum transfers, so that the preference is for $|L|$ to be as low as possible.  Considering the geometry of the laser enclosure, accessible rotation angle, $s_1$, of the laser+cryostat assembly is restricted to $-45^\circ < s_1 < +25^\circ$, avoiding interference of the laser enclosure with upstream incident neutron beam on the positive angle side and with either the downstream neutron beam or the scattered beam on the negative angle side. This $s1$ restriction in turn limits the accessible regions of reciprocal space, which limits $L \in [1,4]$ with an energy transfer range of -2 meV $< E <$ 2 meV when using an incident neutron energy, $E_i$, of 5 meV.

\section{Firmware \& software}

Meanwhile, relevant firmware support modules have been implemented at HYSPEC to synchronize laser pulses with neutron pulses. An analog-to-digital readout card (ADCROC) is utilized in this experiment to acquire meta-data and to trigger the laser synchronously with the SNS accelerator pulsing.  This is the same class of ADCROC reported previously\cite{Granroth:ut5002} and which is already in use at several other beam lines throughout the SNS, but in this use-case the firmware was customized and the graphical user interface in Control System Studio\cite{XGeng2013FirstEB} was tailored to the laser pump/probe setup. A simplified diagram of the control and data acquisition setup is shown in Fig.~\ref{fig:firmware}(b).

The ADCROC acquires meta-data in the same single event as neutrons, which allows processing of the collected meta-data with the same software toolset as for SNS neutron data.  In this experiment, one digital channel is configured as output to control external equipment with a low-voltage time-to-live (LVTTL) pulse signal. A sequence of laser pulses is triggered by LVTTL pulses from the ADCROC that is synchronized to the spallation neutron events with a tunable repetition frequency and delay time as shown in Fig.~\ref{fig:firmware}(a). Two digital channels were configured as inputs to acquire the accelerator time reference and laser trigger reference signals. The relative difference between the accelerator time and laser trigger references sets the delay between laser and neutron pulses arriving at the sample. One input analog channel is used to acquire the laser trigger signal. In order to reduce amount of data acquired by ADCROC, a variable sampling rate functionality is utilized. The data acquisition rate increases when it detects a fast change in signals.

The data reduction is performed using Mantid software,\cite{arnold_mantiddata_2014} similar to other pump-probe workflows.\cite{fancher_time_2018, Peterson_advances_2018}  When the timing between two laser pulse sequences is 60 Hz (matching the spallation frequency), the data may be reduced the same way as for equilibrium measurements.  When the timing between two pulse sequences spans several spallation periods, the data acquired for each frame needs to be reduced separately in order to evaluate against possible decay between two laser pulse sequences.  This is achieved by identifying coincidence of the LVTTL pulse for starting a laser pulse sequence with the initial trigger, and then generating a step log for filtering neutron event data. 

\section{Nonequilibrium structure factors}

\begin{figure*}
\includegraphics[scale = 0.38]{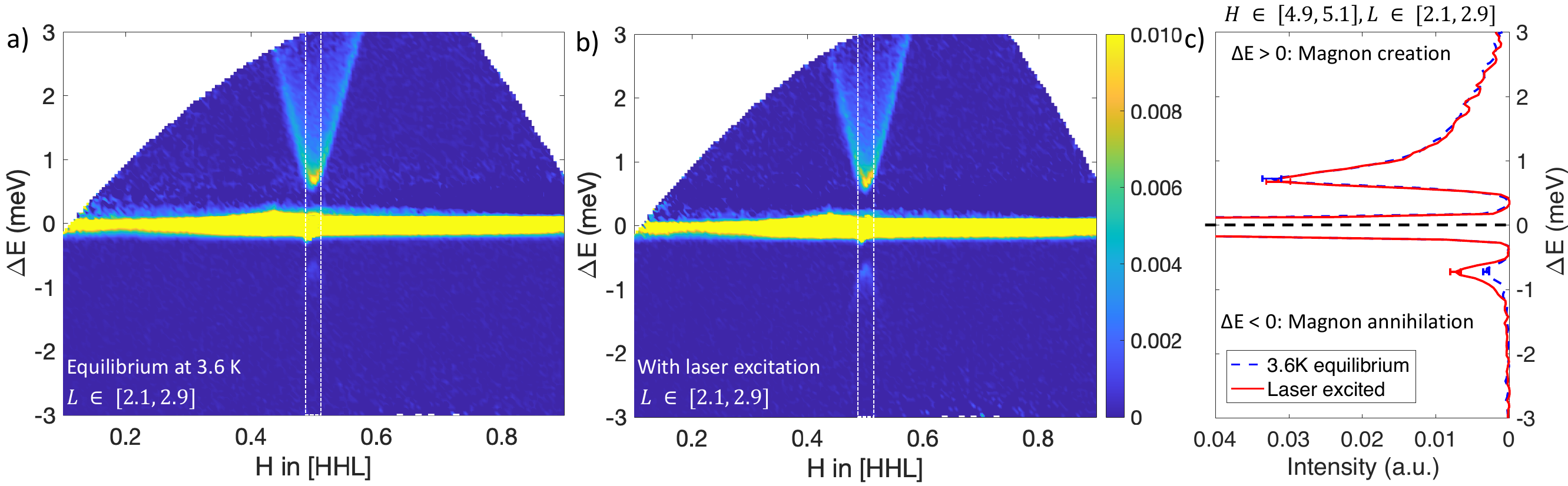}
\caption{Measured intensity $I(\mathbf{Q},\omega)$ of the dynamic structure factor of \rmf ~using HYSPEC (a) with laser excitation (10 pulses per neutron event with no delay time) and (b) at an equilibrium temperature of 3.6K. (c) Comparison between the intensity with laser excitation (solid red line) and at equilibrium (dashed blue line) integrated over a short range of H  around zone center indicated by the dashed boxes in (a) and (b). The intensity on the magnon creation side is well described by the equilibrium data while a photoexcited excessive magnon population around the zone center was observed on the magnon annihilation side.}
\label{fig:spectra}
\end{figure*}

One signature of excess nonequilibrium magnon states is that the ratio of the measured intensity of the dynamic structure factor $I(\mathbf{Q},\omega)$ between the magnon creation and annihilation sides will not follow the principle of detailed balance. At thermodynamic equilibrium, principle of detailed balance has to be strictly following: on the magnon creation side ($\Delta E >$  0), the intensity scales as n$^0_{BE}$(T)+1 where n$^0_{BE}$(T) is the Bose-Einstein distribution and $T$ is the temperature, while it scales as n$^0_{BE}$(T) on the magnon annihilation side ($\Delta E <$ 0). For example, if the laser simply creates a heating effect on the sample and induce an equilibrium state at a raised temperature, the intensity of $I(\mathbf{Q},\omega)$ from both the creation and annihilation sides would increase by the same amount from that at the base temperature. 

The preliminary neutron data from this work has shown intriguing results with a set of rich physics to be explored. Figures~\ref{fig:spectra}a \& b show the measured $I(\mathbf{Q},\omega)$ of \rmf ~ along the H direction with laser excitation (10 pulses per neutron event with no delay time at a grating period of 1 $\mu$m) and in equilibrium at 3.6 K. Since magnon transport is confined to $ab$ plane, no dispersion is observed along the L direction. The measured $I(\mathbf{Q},\omega)$ is integrated between $L \in [2.1,2.9]$ to get the two-dimensional intensity map as shown in Fig.~\ref{fig:spectra}a \& b. By integrating about a small $H$ range around the zone center, Fig.~\ref{fig:spectra}c shows that the intensity on the magnon creation side is well described by the equilibrium data while a photoexcited excessive intensity around the zone center was observed only on the magnon annihilation side. 

\begin{figure*}
\includegraphics[scale = 0.45]{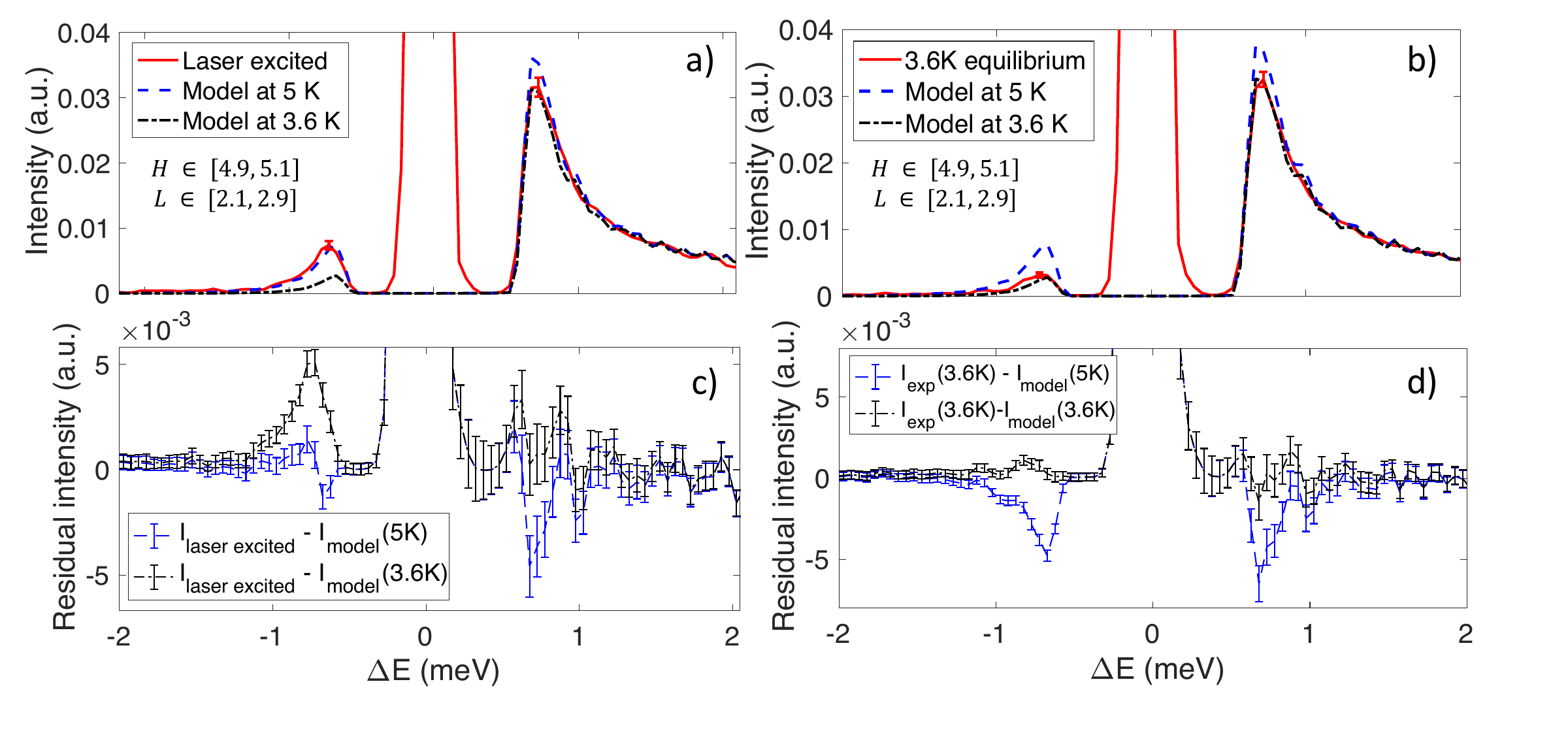}
\caption{The measured intensities (solid red lines) (a) at an equilibrium temperature of 3.6 K and (b) with laser excitation are compared with the modeled intensity at both 3.6 K (dot-dashed black lines) and 5 K (dashed blue lines).  The residual intensity as a function of energy transfer $\Delta E$ is calculated by subtracting the measured intensity (c) with laser excitation and (d) at an equilibrium temperatures with the modeled intensity at both 3.6 K (dot-dashed black lines) and 5K (dashed blue lines).}
\label{fig:Residue}
\end{figure*}

To show this increase in intensity on the magnon annihilation side is non-thermal, we calculated the intensity of the total dynamic structure factor of \rmf~at thermodynamic equilibrium given as
\begin{widetext}
\begin{eqnarray}
&&I(\mathbf{Q},\omega) = \int d\mathbf{Q'}d\omega' F(\mathbf{Q'})^2 S(\mathbf{Q'},\omega')R(\mathbf{Q}-\mathbf{Q'},\omega-\omega') \nonumber \\ 
&&\times \sum_\mathbf{Q'}\left\{\delta(\hbar\omega'-\hbar\omega_{Q'})[n^0_{BE}(\omega_{Q'},T)+1] + \delta(\hbar\omega'+\hbar\omega_{Q'})n^0_{BE}(\omega_{Q'},T)\right\} 
\label{eq:I}
\end{eqnarray}
\end{widetext}
where $F(\mathbf{Q})$ is the Mn$^{2+}$ magnetic form factor and $S(\mathbf{Q},\omega)$ is the dynamic structure factor of \rmf~given in Ref.~\cite{huberman_twomagnon_2005}. HYSPEC instrumental resolution function, $R(\mathbf{Q},\omega)$ in Eq.~(\ref{eq:I}), has been modeled employing the software MCViNE.\cite{lin_mcvine_2016} Three separate Monte Carlo simulations have been performed to estimate the contributions to the resolution given by the instrument optics, the sample, and the detectors. Specifically, we first perform a beam simulation which simulates the neutron interactions with all the optics components present along the beam path such as neutron guides, choppers, slits, and the monochromator.  This simulation has been performed with a statistic of 10 million neutrons at the source, to achieve good statistical average. A kernel containing the information about the sample shape, size, orientations along the axes, lattice constants, and UB matrix, is then generated for the second part of the Monte Carlo. MCViNE loads the beam simulation, and it simulates the scattering with the sample which is rotated within the same angular range used during the measurement. Finally, the scattered neutron trajectories are recorded on the detector array, reduced, and placed on a regular grid within the same momentum and energy transfer ranges used for the data. The resolution ellipsoids in the four dimensional space are then simulated, fitted, and the resulting FWHMs saved in the final output file. Once this procedure is over, a convolution calculation of the resolution function with the predicted dynamic structure factor represented by Eq.~(\ref{eq:I}) is performed to get the final intensity along the desired Q-direction.  

Figures~\ref{fig:Residue}(a) \& (b) compare the measured intensity with laser excitation and at 3.6 K equilibrium with the modeled intensity at both 3.6 K and 5 K. While the measured intensity at equilibrium is described well by the 3.6 K model on both sides, the annihilation and creation intensity with laser excitation has to be described by two different temperatures. Figures~\ref{fig:Residue}(c) \& (d) give the residual intensity when the measured intensity is subtracted by the modeled intensity at both 3.6 K and 5K. At an equilibrium temperature of 3.6 K, the residual intensity is around zero on both creation and annihilation sides within the experimental uncertainty. The residual intensity with respect to the modeled intensity at 5K shows negative peaks on both sides, indicating more magnon states will be populated according to Bose-Einstein distribution and can be created and annihilated by neutrons. However, with laser excitation as shown in Fig.~\ref{fig:Residue}(c), the residual intensity is around zero only on the annihilation side with respect to the modeled intensity at 5K while the negative peak on the creation side is still present. This observation excludes the observed excess intensity on the annihilation side with laser excitation is simply due to thermal effects. 

Since the effective temperatures are different on the creation and annihilation sides, the principle of detailed balance is clearly broken under laser excitation. The excess intensity on the annihilation side comes from extra magnon population created by laser excitations. Because the intensity on the creation side still follows the equilibrium distribution at 3.6 K, this suggests the bulk temperature of the crystal is still at 3.6 K . The observed excessive magnon population cannot be described by the equilibrium distribution at the bulk temperature, a key indication that these magnons are at out-of-equilibrium states. A complete set of out-of-equilibrium INS measurements and detailed discussion on the physical origins of these out-of-equilibrium states and the effects of grating period on the magnon re-equilibration processes will be presented elsewhere. 

\section{Conclusion}

This work enabled and demonstrated the detection of non-equilibrium magnon states in a 2D quantum magnet using a laser excitation and a neutron probe. So far, this proof-of-principle demonstration detected a few snapshots of non-equilibrium states in one model system, and opens an entirely new research direction to probe and understand time-resolved non-equilibrium processes of matter with microscopic momentum resolution using neutrons. It is worth emphasizing that each material system poses unique technical challenges and should not be limited to the exact sample environment described above. Optical cryostats other than CRYO-B, lasers with different power, wavelength, and pulse duration, and even neutron beamlines other than HYSPEC will be used as needed to probe the relevant physics and quantum material systems of interest. In this sense, this work will serve as a prototype for future pump-probe INS measurements.

The laser-neutron pump-probe technique developed in this work will have spin-off potential for studying a broader range of phenomena using neutrons under non-equilibrium conditions. Out-of-equilibrium phonon dynamics, structural and magnetic phase transitions induced by high-power pulsed laser excitation, and diffuse scattering caused by charge density wave melting due to laser heating could be potentially studied using a combination of laser excitation and neutron probes. Moreover, the work is particularly timely as the Second Target Station (STS) will prove even more effective at measuring out-of-equilibrium dynamics using neutrons.\cite{Sala_CHESS_2022, Galambos2015TechnicalDR} STS will feature high brightness cold neutrons that are suited for experiments with smaller samples than currently possible, meaning more material systems with weak scattering will be compatible with the proposed laser-pump neutron-probe spectroscopy. The Proton Power Upgrade at SNS will also enable a new structure in neutron pulse sequences, providing additional flexibility in time delay scheme between laser and neutron pulses. With the promise of new, more powerful neutron sources in the future, the possibilities for time-resolved neutron scattering experiments will improve and are bound to garner interest.

\section*{Acknowledgements}

The authors would like to acknowledge Chris Redmon for CRYO-B modifications, David Connor for mechanical mounting and alignment, and Mariano Ruiz-Rodriguez for setting up the graphical user interface that manages ADCROC timing settings.

\section*{DATA AVAILABLITY}

The data that support the findings of this study are available from the corresponding author upon reasonable request.

%\bibliography{MyRef.bib}


\end{document}